\documentclass[11pt]{article}
\usepackage{moriond,epsfig}

\def\Journal#1#2#3#4{{#1} {\bf #2}, #3 (#4)}

\def\EPJ{{\em Eur. Phys. J.} C}
\def\NCA{\em Nuovo Cimento}

\def\NPB{{\em Nucl. Phys.} B}
\def\NPBPS{{\em Nucl. Phys.} B (Proc.\ Suppl.)}
\def\RMP{\em Rev. Mod. Phys.}
\def\PL{\em Phys. Lett.}
\def\PLB{{\em Phys. Lett.}  B}
\def\PRt{\em Phys. Rept.}
\def\PRL{\em Phys. Rev. Lett.}
\def\PRD{{\em Phys. Rev.} D}
\def\PR{\em Phys. Rev.}

\def\JETP{\em J. Exp. Theor. Phys.}
\def\JETPL{\em J. Exp. Theor. Phys. Lett.}
\def\SJNP{\em Sov. J. Nucl. Phys.}

\def\ARNPS{\em Annu.\ Rev.\ Nucl.\ Part.\ Sci.}

\def\be{\begin{equation}}
\def\ee{\end{equation}}
\def\bea{\begin{eqnarray}}
\def\eea{\end{eqnarray}}

\def\etal{{\it et al.}}
\begin{document}

\title{\vspace*{-1.cm}
\begin{flushright}
{\normalsize\bf LAL 07-88}\\
\vspace*{-0.1cm}
{\normalsize November 2007}
\end{flushright}
\vspace*{3cm}
\LARGE Muon $g-2$: a mini review}

\author{\Large\bf Zhiqing Zhang\vspace*{0,5cm} }

\address{LAL, Univ Paris-Sud, CNRS/IN2P3, Orsay, France}

\maketitle\abstracts{\normalsize
The current status of the experimental measurements and 
theoretical predictions 
of the anomalous magnetic moment of the muon $a_\mu$ is briefly reviewed. 
The emphasis is put on the evaluation of the hadronic contribution to $a_\mu$
as it has the largest uncertainty among all Standard Model contributions.
The precision of the hadronic contribution is driven by the input $e^+e^-$ data
predominantly from the $\pi^+\pi^-$ channel.
Including the latest experimental data on $e^+e^-$ annihilation into hadrons
from CMD2 and SND for the $\pi^+\pi^-$ channel and {\sc BaBar} for multihadron
final states, the updated Standard Model prediction disagrees with 
the measurement dominated by BNL by $3.3$ standard 
deviations, with the theoretical precision exceeding
the experimental one.}

\section{Introduction}

For a charged elementary particle with $1/2$ intrinsic spin such as muon, 
its magnetic dipole moment $\vec{\mu}$ is aligned with its spin 
$\vec{s}$ as:
\begin{equation}
\vec{\mu}=g\left(\frac{q}{2m}\right)\vec{s}\,,
\end{equation}
where $q=\pm e$ is the charge of the particle in unit of the electron charge
and $g$ is the gyromagnetic ratio. In the classic Dirac theory, $g=2$.
In the Standard Model (SM), quantum loop effects induce a small correction,
which is quantified by $a_\mu=(g_\mu-2)/2$, the so-called anomalous magnetic 
moment or the magnetic anomaly.

There has been a long history in measuring and calculating $a_\mu$. 
In particular the steadily improving precision of both the measurements and 
the predictions of $a_\mu$ and the disagreement observed between the two 
have led the study of $a_\mu$ one of the most active research fields in 
particle physics in recent years.


The paper is organized as follows. In section \ref{sec:meas}, the measurement
history and the current world average value of $a_\mu$ are presented.
In section \ref{sec:th}, different components of the SM contributions to
$a_\mu$ are reviewed. Section \ref{sec:dis} is reserved for discussions 
followed by conclusion and prospects in section \ref{sec:end}.

\section{The Measurement of $a_\mu$}\label{sec:meas}

A compilation of the major experimental efforts in measuring $a_\mu$ over
the last five decades is given in Table~\ref{tab:meas} (a modified version
of Table 1 from a recent review article~\cite{mdr07}). 
Starting from the experiment at the Columbia-Nevis cyclotron,
where the spin rotation of a muon in a magnetic field was observed for
the first time, the experimental precision of $a_\mu$ has seen
constant improvement first through three experiments at CERN in the
sixties and seventies and more recently with E821 at the Brookhaven 
National Laboratory (BNL). The current world average value reaches 
a relative precision of $0.54\,{\rm ppm}$ 
(parts per million).
\begin{table}[ht]
\caption{Measurements of the muon magnetic anomaly $a_\mu$, where the
value in parentheses stands for either the total experimental error or 
the statistical and systematic ones.
\label{tab:meas}}
\vspace{0.4cm}
\begin{center}
\begin{tabular}{|l|c|l|r|}
\hline
Experiment & Beam & Measurement & $\delta a_\mu/a_\mu$ \\ \hline
Columbia-Nevis($1957$)~\cite{nevis57} & $\mu^+$ & $g=2.00\pm 0.10$ & \\
Columbia-Nevis($1959$)~\cite{nevis59} & $\mu^+$ & $0.001\,13^{+(16)}_{-(12)}$ 
 & $12.4\%$ \\ \hline
CERN $1$($1961$)~\cite{cern1-61} & $\mu^+$ & $0.001\,145(22)$ & $1.9\%$ \\
CERN $1$($1962$)~\cite{cern1-62} & $\mu^+$ & $0.001\,162(5)$ & $0.43\%$ \\
CERN $2$($1968$)~\cite{cern2-68} & $\mu^\pm$ & $0.001\,166\,16(31)$ & 
 $265\,$ppm \\
CERN $3$(1975)~\cite{cern3-75} & $\mu^\pm$ & 0.001\,165\,895(27) & 23\,ppm \\
CERN 3(1979)~\cite{cern3-79} & $\mu^\pm$ & 0.001\,165\,911(11) & 7.3\,ppm \\
\hline
BNL E821(2000)~\cite{bnl00} & $\mu^+$ & 0.001\,165\,919\,1(59) & 5\,ppm \\
BNL E821(2001)~\cite{bnl01} & $\mu^+$ & 0.001\,165\,920\,2(16) & 1.3\,ppm \\
BNL E821(2002)~\cite{bnl02} & $\mu^+$ & 0.001\,165\,920\,3(8) & 0.7\,ppm \\
BNL E821(2004)~\cite{bnl04} & $\mu^-$ & 0.001\,165\,921\,4(8)(3) & 0.7\,ppm \\
 \hline
World Average(2004)~\cite{bnl04,wa04} & $\mu^\pm$ & 0.001\,165\,920\,80(63) & 
 0.54\,ppm \\ \hline
\end{tabular}
\end{center}
\end{table}

The muon magnetic anomaly $a_\mu$ in all modern experiments is determined
by the following method.
For an ensemble of polarized muons which are moving in a storage ring
in a highly uniform magnetic field $\vec{B}$ (perpendicular to muon spin and 
orbit plane) and a vertically focusing quadrupole field $\vec{E}$,
the frequency difference $\omega_a$ between the spin procession $\omega_s$
and the cyclotron motion $\omega_c$ is described by
\be
\vec{\omega}_a\equiv \vec{\omega}_s-\vec{\omega}_c=\frac{e}{m_\mu c}
\left[ a_\mu\vec{B}-\left(a_\mu-\frac{1}{\gamma^2-1}\right)(\vec{\beta}\times\vec{E})\right]\,, \hspace{5mm}{\rm when}\hspace{2mm} \vec{B}\cdot\vec{\beta}=\vec{E}\cdot\vec{\beta}=0
\ee
where $\vec{\beta}$ represents the muon direction. The second term in
parentheses vanishes at $\gamma=29.3$ (the magic momentum) and 
the electrostatic focusing does not affect the spin.
The key to the experiment is to determine frequency $\omega_a$ to high
precision and to measure the average magnetic field to equal or
better precision.

In comparison with the electron magnetic anomaly,
$a_e$ is more precisely measured~\cite{ae_mea} ($0.7\,{\rm ppb}$),
but $a_\mu$ is more sensitive to new physics
effects by about $m^2_\mu/m^2_e\simeq40\,000$ because of its large mass value.

\section{Prediction of the Standard Model Contributions}\label{sec:th}

\begin{figure}[h]
   \newcommand\thisSize{2.8cm}
   \newcommand\thisHspace{\hspace{0.9cm}}
   \centerline{\epsfxsize\thisSize\epsffile{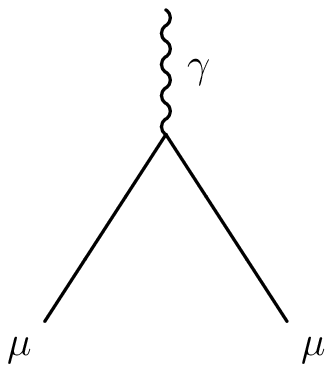}
               \thisHspace
               \epsfxsize\thisSize\epsffile{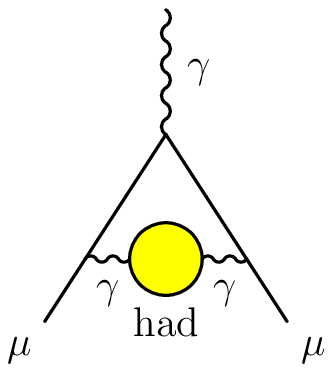}
               \thisHspace
               \epsfxsize\thisSize\epsffile{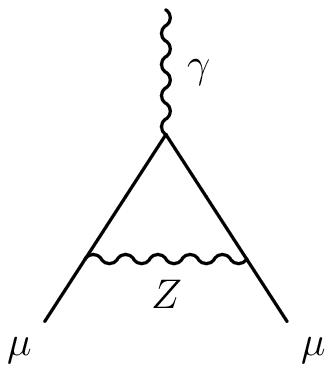}
               \thisHspace
               \epsfxsize\thisSize\epsffile{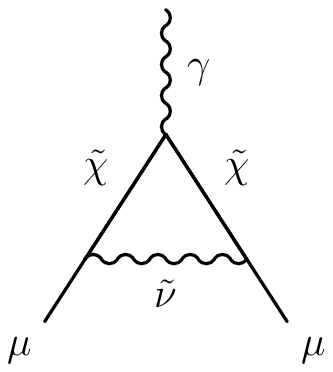}}
   \vspace{0.0cm}
   \centerline{\epsfxsize\thisSize\epsffile{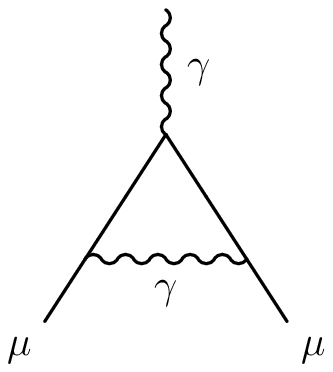}
               \thisHspace
               \epsfxsize\thisSize\epsffile{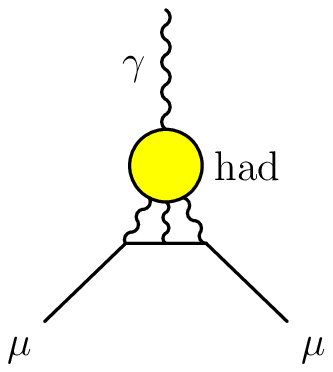}
               \thisHspace
               \epsfxsize\thisSize\epsffile{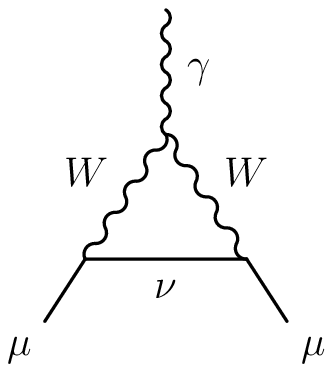}
               \thisHspace
               \epsfxsize\thisSize\epsffile{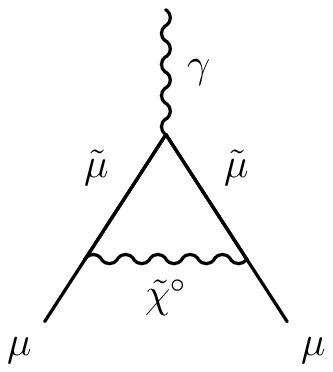}}               
   \vspace{0.3cm}
   \caption{Representative diagrams contributing to $a_\mu$.
        First column: lowest-order diagram (upper) and first
        order QED correction (lower); second column: lowest-order 
        hadronic contribution (upper) and hadronic 
        light-by-light scattering (lower); third column: weak
        interaction diagrams; last column: possible 
        contributions from lowest-order Supersymmetry.}
   \label{fig:amu}
\end{figure}

In the SM, the muon magnetic anomaly $a_\mu$ receives contributions from
all electromagnetic (QED), weak and strong (hadronic) sectors and can be 
conveniently written as:
\begin{equation}
a_\mu^{\rm SM}=a_\mu^{\rm QED}+a_\mu^{\rm weak}+a_\mu^{\rm had}\,.
\label{eq:amu}
\end{equation}
Their representative diagrams are shown in Fig.~\ref{fig:amu}, which also includes two example contributions from new
particles in supersymmetry models. Thus comparison of the precision 
measurement and theory tests the validity of the SM at its quantum loop level 
and probes effects of new physics.

\subsection{QED and Weak Contributions}

The QED correction, which includes all photonic and leptonic ($e$, $\mu$ and
$\tau$) loops, is by far the dominant contribution in the SM:
\bea
a^{\rm QED}_\mu=\hspace{-5mm}&&\frac{\alpha}{2\pi}+
0.765\,857\,410(27)\left(\frac{\alpha}{\pi}\right)^2+
24.050\,509\,64(43)\left(\frac{\alpha}{\pi}\right)^3+
130.9916(80)\left(\frac{\alpha}{\pi}\right)^4+\nonumber \\
&&+663(20)\left(\frac{\alpha}{\pi}\right)^5+\cdots\,,
\eea
where the lowest-order Schwinger term ($\alpha/2\pi$) was known since 
1948~\cite{js48}, the coefficients are analytically known for terms up to 
$(\alpha/\pi)^3$, numerically calculated for the fourth term 
and recently estimated for the fifth term~\cite{kn06}. 
Using $\alpha$ extracted from
the latest $a_e$ measurement~\cite{ae_mea}, one has
\be
a^{\rm QED}_\mu=116\,584\,718.09(0.14)_{\rm 5th\,order}(0.08)_{\delta\alpha}\times 10^{-11}\,.
\ee

The week contributions, involving heavy $Z$, $W^\pm$ or Higgs particles,
are suppressed by at least a factor 
$\frac{\alpha}{\pi}\frac{m^2_\mu}{M^2_W}\simeq 4\times 10^{-9}$.
At one-loop order,
\begin{eqnarray}
a^{\rm week}_\mu[\mbox{1-loop}]\!&=\!&\frac{G_\mu m^2_\mu}{8\sqrt{2}\pi^2}
\left[\frac{5}{3}+\frac{1}{3}\left(1-4\sin^2\!\theta_W\right)^2+
{\cal O}\left(\frac{m^2_\mu}{M^2_W}\right)+
{\cal O}\left(\frac{m^2_\mu}{M^2_H}\right)\right]\\
\!&=\!&194.8\times 10^{-11}\,, \hspace{1cm} {\rm for}\hspace{1mm}
\sin^2\!\theta_W\equiv 1-\frac{M^2_W}{M^2_Z}\simeq0.223\,.
\end{eqnarray}
Two-loop corrections are relatively large and negative
\be
a^{\rm weak}_\mu[\mbox{2-loop}]=-40.7(1.0)(1.8)\times 10^{-11}\,,
\ee
where the errors stem from quark triangle loops and the assumed Higgs mass
range $M_H=150^{+100}_{-40}\,{\rm GeV}$. The three-loop leading logarithms are
negligible, ${\cal O}(10^{-12})$, implying in total
\be
a^{\rm weak}_\mu=154(1)(2)\times 10^{-11}\,.
\ee

\subsection{Hadronic Contributions}

The hadronic contributions are associated with quark and gluon loops. 
They cannot be calculated from first principles because of the low energy 
scale involved.
Fortunately, owing to unitarity and to the analyticity of the 
vacuum polarization function, the lowest-order hadronic vacuum polarization 
contribution to $a_\mu$ can be computed via the dispersion 
integral~\cite{gr69} using the ratio $R^{(0)}(s)$ of the bare cross 
section~\footnote{The bare cross section is defined
as the measured cross section corrected for initial state radiation, electron
vertex contributions and vacuum polarization effects in the photon propagator
but with photon radiation in the final state included~\cite{dhz06}.}
for $e^+e^-$ annihilation into hadrons to the pointlike muon pair cross
section at center-of-mass energy $\sqrt{s}$
\be
a^{\rm had,LO}_\mu=\frac{1}{3}\left(\frac{\alpha}{\pi}\right)^2
\int^\infty_{m^2_\pi}ds\frac{K(s)}{s}R^{(0)}(s)\,,
\ee
where $K(s)$ is the QED kernel~\cite{br68} $K(s)=x^2\left(1-\frac{x^2}{2}\right)+(1+x)^2\left(1+\frac{1}{x^2}\right)\left[\ln(1+x)-x+\frac{x^2}{2}\right]+x^2\ln\! x\frac{1+x}{1-x}$, with $x=\frac{1-\beta_\mu}{1+\beta_\mu}$ and $\beta_\mu=\left(1-\frac{4m^2_\mu}{s}\right)^{1/2}$. 
The kernel function $K(s)\sim \frac{1}{s}$ gives weight to the low energy part
of the integral. About $91\%$ of the total contribution to $a^{\rm had,LO}_\mu$ is accumulated at $\sqrt{s}$ below $1.8\,{\rm GeV}$ and $73\%$ of
$a^{\rm had,LO}_\mu$ is covered by the $\pi\pi$ final state, 
which is dominated by the $\rho (770)$ resonance.


A detailed compilation of all the experimental data used in the evaluation
of the dispersion integral prior to 2004 is provided in 
Refs.~\cite{dehz03a,dehz03b}. Since then, a few precise measurements
have been published. A list of experiments for the dominant $\pi\pi$
channel is shown in Table~\ref{tab:2pi}.

\begin{table}[h]
\caption{A list of measurements of $e^+e^-$ annihilation into hadrons in
the $\pi^+\pi^-(\gamma)$ channel.\label{tab:2pi}}
\vspace{0.4cm}
\begin{center}
\begin{tabular}{|l|c|c|c|c|}
\hline
Experiment & $N_{\rm data}$ & Energy range (GeV) & $\delta$(stat.) & $\delta$(syst.) \\ \hline
DM1 (1978)~\cite{dm1} & $16$ & $0.483-1.096$ & $(6.6-40)\%$ & $2.2\%$ \\
TOF (1981)~\cite{tof} & $4$ & $0.400-0.460$ & $(14-20)\%$ & $5\%$ \\
OLYA (1979, 1985)~\cite{olya79,olya-cmd85} & $2+77$ & $0.400-1.397$ & $(2.3-35)\%$ & $4\%$ \\
CMD (1985)~\cite{olya-cmd85} & $24$ & $0.360-0.820$ & $(4.1-10.8)\%$ & $2\%$ \\
DM2 (1989)~\cite{dm2} & $17$ & $1.350-2.215$ & $(17.6-100)\%$ & $12\%$ \\ 
CMD2 (2003)~\cite{cmd2-03} & $43$ & $0.611-0.962$ & $(1.8-14.1)\%$ & $0.6\%$ \\
KLOE (2005)~\cite{kloe} & $60$ & $0.600-0.970$ & $(0.5-2.1)\%$ & $(1.2-3.8)\%$ \\
SND (2006)~\cite{snd} & $45$ & $0.390-0.970$ & $(0.5-2.1)\%$ & $(1.2-3.8)\%$ \\
CMD2$_{\rm low}$ (2006)~\cite{cmd2low} & $10$ & $0.370-0.520$ & $(4.5-7)\%$ & $0.7\%$ \\
CMD2$_{\rm rho}$ (2006)~\cite{cmd2rho} & $29$ & $0.600-0.970$ & $(0.5-4.1)\%$ & $0.8\%$ \\
CMD2$_{\rm high}$ (2006)~\cite{cmd2high} & $36$ & $0.980-1.380$ & $(4.5-18.4)\%$ & $(1.2-4.2)\%$ \\ \hline
\end{tabular}
\end{center}
\end{table}

\begin{figure}[!h]
\begin{center}
\psfig{figure=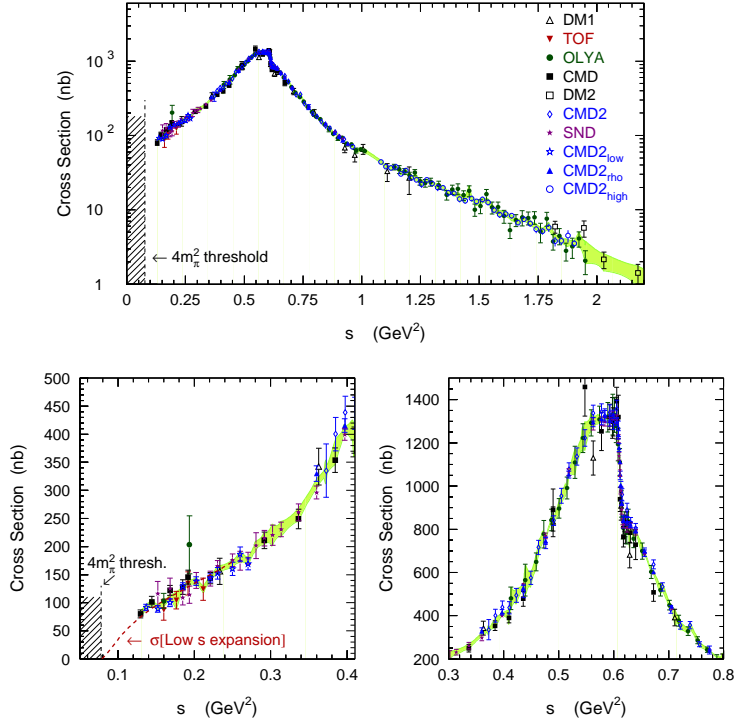,height=10cm}
\end{center}
\vspace{-5mm}
\caption{Comparison of $\pi^+\pi^-$ spectral functions expressed as $e^+e^-$ cross sections. The band corresponds to combined data used in the numerical integration.\label{fig:2pi}}
\end{figure}

The $\pi\pi$ data are compared in Fig.~\ref{fig:2pi}. Closer inspections show that the most precise measurements from the 
annihilation experiments SND and CMD2 at Novosibirsk
are in good agreement. They differ however in shape with those measured by 
KLOE using the radiative return method at DA$\Phi$NE~\cite{kloe} (see Sec.~\ref{sec:tau-ee}). 
Before this is clarified, the KLOE data are
not used in some of the recent evaluations of $a^{\rm had,LO}_\mu$.

\begin{figure}[!h]
\vspace*{-1.3cm}
\begin{center}
\psfig{figure=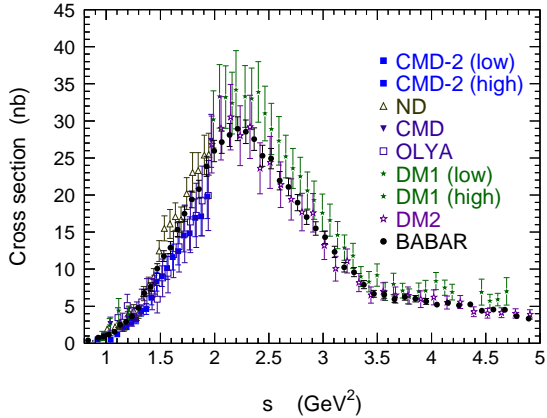,height=7.4cm}
\end{center}
\vspace{-7mm}
\caption{The measured cross section for $e^+e^-\rightarrow 2\pi^+2\pi^-$ from
{\sc BaBar} compared to previous measurements.\label{fig:4pi}}
\end{figure}

In addition to the dominant $\pi\pi$ mode, results from the {\sc BaBar} 
experiments are being produced on multihadron final states using also
radiative return~\cite{babar}. Benefiting from its big initial 
center-of-mass energy of $10.6\,{\rm GeV}$, hard-radiated 
photon detected at large angle and high statistics data sample, 
the {\sc BaBar} measurements are precise 
over the whole mass range. One example is shown in Fig.~\ref{fig:4pi} 
in comparison with earlier measurements.

Including these new input $e^+e^-$ data, a preliminary update~\footnote{It is
preliminary as some of the new $e^+e^-$ data used were still in their 
preliminary form.} of
$a^{\rm had,LO}_\mu$ is performed~\cite{dehz06}
and shown in Table~\ref{tab:amures}.
\begin{table}[htb]
\caption{Summary of the $a^{\rm had,LO}_\mu$ contributions from $e^+e^-$
annihilation and $\tau$ decays.\label{tab:amures}}
\vspace{0.4cm}
\begin{center}
\begin{tabular}{|l|c|c|c|}
\hline
Modes & Energy [GeV] & $a^{\rm had,LO}_\mu(e^+e^-) [10^{-10}]$ & $a^{\rm had,LO}_\mu(\tau) [10^{-10}]$ \\ \hline
Low $s$ expansion & $2m_\pi-0.5$ & $55.6\pm 0.8\pm 0.1_{\rm rad}$ & $56.0\pm 1.6\pm 0.3_{\rm SU(2)}$ \\
$\pi^+\pi^-$ & $0.5-1.8$ & $449.0\pm 3.0\pm 0.9_{\rm rad}$ & $464.0\pm 3.0\pm 2.3_{\rm SU(2)}$ \\
$\pi^+\pi^-2\pi^0$ & $2m_\pi-1.8$ & $16.8\pm 1.3\pm 0.2_{\rm rad}$ & $21.4\pm 1.3\pm 0.6_{\rm SU(2)}$ \\
$2\pi^+2\pi^-$ & $2m_\pi-1.8$ & $13.1\pm 0.4\pm 0.0_{\rm rad}$ & $12.3\pm 1.0\pm 0.4_{\rm SU(2)}$ \\
$\omega(782)$ & $0.3-0.81$ & $38.0\pm 1.0\pm 0.3_{\rm rad}$ & $-$ \\
$\phi(1020)$ & $1.0-1.055$ & $35.7\pm 0.8\pm 0.2_{\rm rad}$ & $-$ \\
Other excl. & $2m_\pi-1.8$ & $24.3\pm 1.3\pm 0.2_{\rm rad}$ & $-$ \\
$J/\psi, \psi(2S)$ & $3.08-3.11$ & $7.4\pm 0.4\pm 0.0_{\rm rad}$ & $-$ \\
$R$ [QCD] & $1.8-3.7$ & $33.9\pm 0.5_{\rm QCD}$ & $-$ \\
$R$ [data] & $3.7-5.0$ & $7.2\pm 0.3\pm 0.0_{\rm rad}$ & $-$ \\
$R$ [QCD] & $5.0-\infty$ & $9.9\pm 0.2_{\rm QCD}$ & $-$ \\ \hline
sum & $2m_\pi-\infty$ & $690.8(3.9)(1.9)_{\rm rad}(0.7)_{\rm QCD}$ & $710.1(5.0)(0.7)_{\rm rad}(2.8)_{\rm SU(2)}$ \\ \hline
\end{tabular}
\end{center}
\end{table}
There is no new tau data since the previous evaluation, therefore the $\tau$ 
based calculation is taken directly from Ref.~\cite{dehz03b}.
The evaluation using $\tau$ data is made~\cite{adh98} 
by relating the vector spectral
functions from $\tau \rightarrow\nu_\tau +$ hadrons decays to isovector 
$e^+e^-\rightarrow$ hadrons cross sections by isospin rotation. 
All known isospin breaking effects are then taken into account~\cite{dehz03a,dehz03b}.



The higher order (NLO) hadronic contributions $a^{\rm had,NLO}_\mu$ involve
one hadronic vacuum polarization insertion with an additional loop (either
photonic or leptonic or another hadronic vacuum polarization). 
They can be evaluated~\cite{krause} with the same $e^+e^-\rightarrow$ hadrons 
data sets used for $a^{\rm had,LO}_\mu$. The numerical value~\cite{hmnt03}
reads
\be
a^{\rm had,NLO}_\mu=-9.79(0.09)_{\rm exp}(0.03)_{\rm rad}\times 10^{-10}
\ee
where the first and second errors correspond respectively to the experimental
uncertainty of the $e^+e^-$ data and the radiative correction uncertainty.

Another higher order hadronic contribution to $a_\mu$ is from the hadronic 
light-by-light scattering (illustrated with the lower figure in the second 
column in Fig.~\ref{fig:amu}). Since it invokes a four-point correlation 
function, a dispersion relation approach using data is not possible at present.
Instead, calculations~\cite{lbl} involving pole insertions (or Goldstone boson
exchanges), short distance quark loops and charged pion (and kaon) loops 
have been individually performed in a large $N_c$ QCD approach.
A representative value used in Ref.~\cite{dehz03a} was 
$a^{\rm had,LBL}_\mu=8.6(3.5)\times 10^{-10}$.

A new analysis~\cite{mv}, which taks into account the proper matching of 
asymptotic short-distance behavior of pseudoscalar and axial-vector 
contributions with the free quark loop behavior, leads to $a^{\rm had,LBL}_\mu=
13.6(2.5)\times 10^{-10}$. However, as pointed out in Ref.~\cite{dm04},
several small but negative contributions such as charged pion loops and
scalar resonances were not included in the latter calculation, thus in
a recent update evaluation~\cite{dehz06} of $a_\mu$, the following value
\be
a^{\rm had,LBL}_\mu=12.0(3.5)\times 10^{-10}
\ee
was used~\footnote{A different evaluation~\cite{hmnt06} used directly 
the value of Ref.~\cite{mv} of $13.6(2.5)\times 10^{-10}$.}. 
This is consistent with the value
$a^{\rm had,LBL}_\mu=11(4)\times 10^{-11}$, suggested in Ref.~\cite{bp07}.
The uncertainty $a^{\rm had,LBL}_\mu$, being the second largest one next to
$a^{\rm had,LO}_\mu$, clearly needs improvement in the near future.

\begin{figure}[htb]
\begin{center}
\psfig{figure=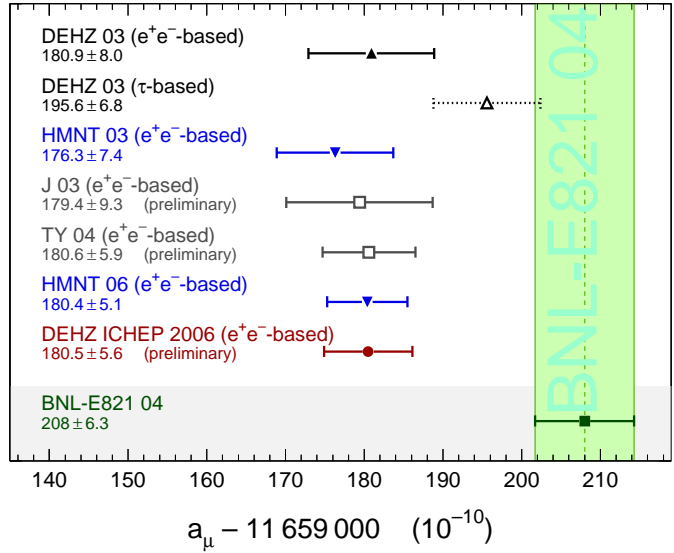,height=7.5cm}
\end{center}
\caption{Comparison of recent theoretical evaluations of $a_\mu$ with
the BNL measurement.\label{fig:amures}}
\end{figure}

Adding all SM contributions together, the comparison from recent 
evaluations~\cite{dehz03b,hmnt03,j03,ty04,hmnt06,dehz06}
with the measurement is shown in Fig.~\ref{fig:amures}.
While the $\tau$ data-based calculation agrees with the measurement within
the errors, the $e^+e^-$ data-based evaluations show a deviation of around
$3.3$ standard deviations.

\section{Discussions}\label{sec:dis}

\subsection{Tau Data versus $e^+e^-$ Data}\label{sec:tau-ee}

The $\tau$ data used in the $a_\mu$ evaluation is the averaged one from
the LEP experiments ALEPH~\cite{aleph} and OPAL~\cite{opal} and the CLEO 
experiment~\cite{cleo}. The data are compared in Ref.~\cite{dhz06} and found
in good agreement in particular for the two most precise data from ALEPH
and CLEO. These data are complementary as the ALEPH data are more precise
below the $\rho$ peak while CLEO has the better precision above.

\begin{figure}[htb]  
  \centerline{
          \psfig{file=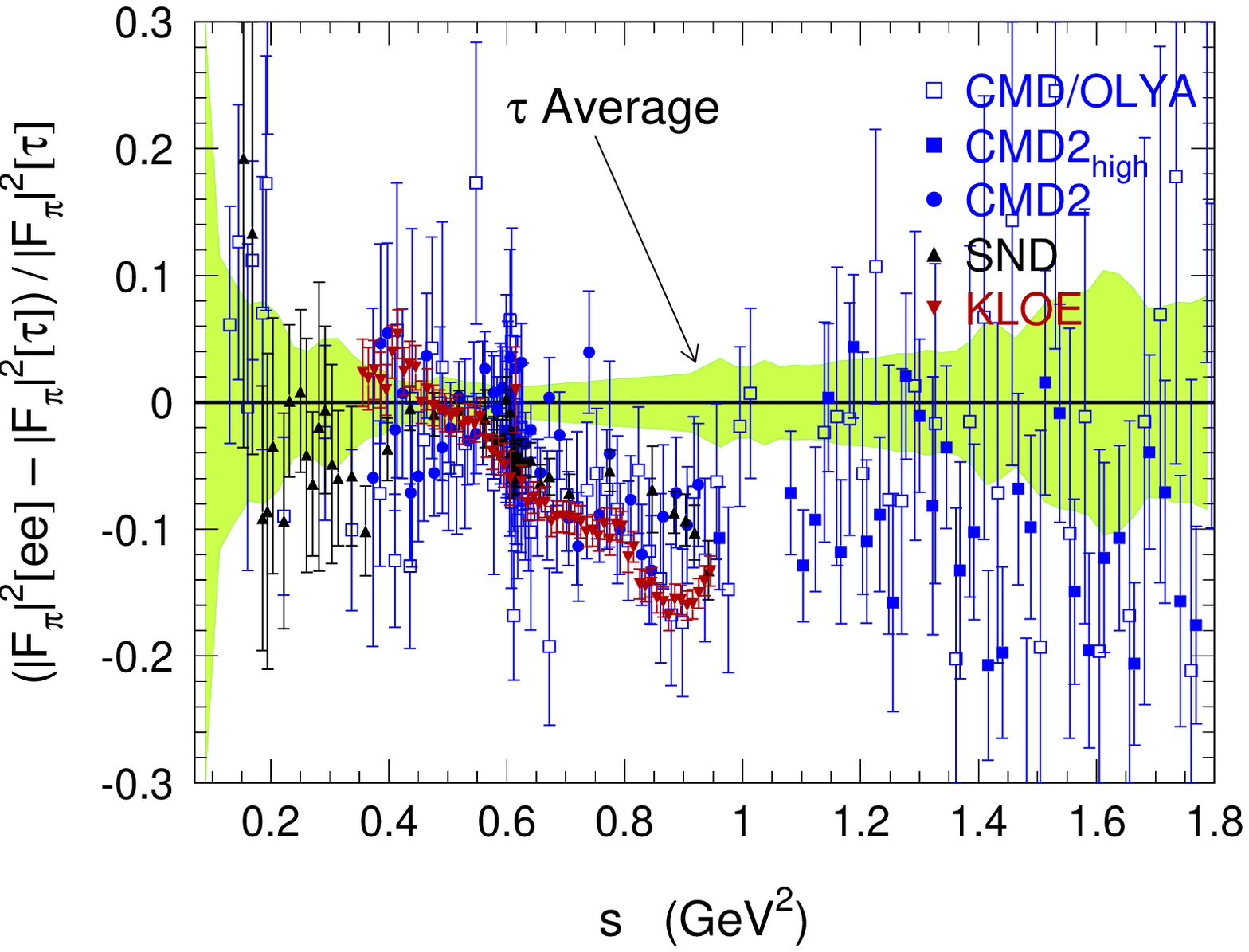,width=78mm}
          \hspace{0.05cm}
          \psfig{file=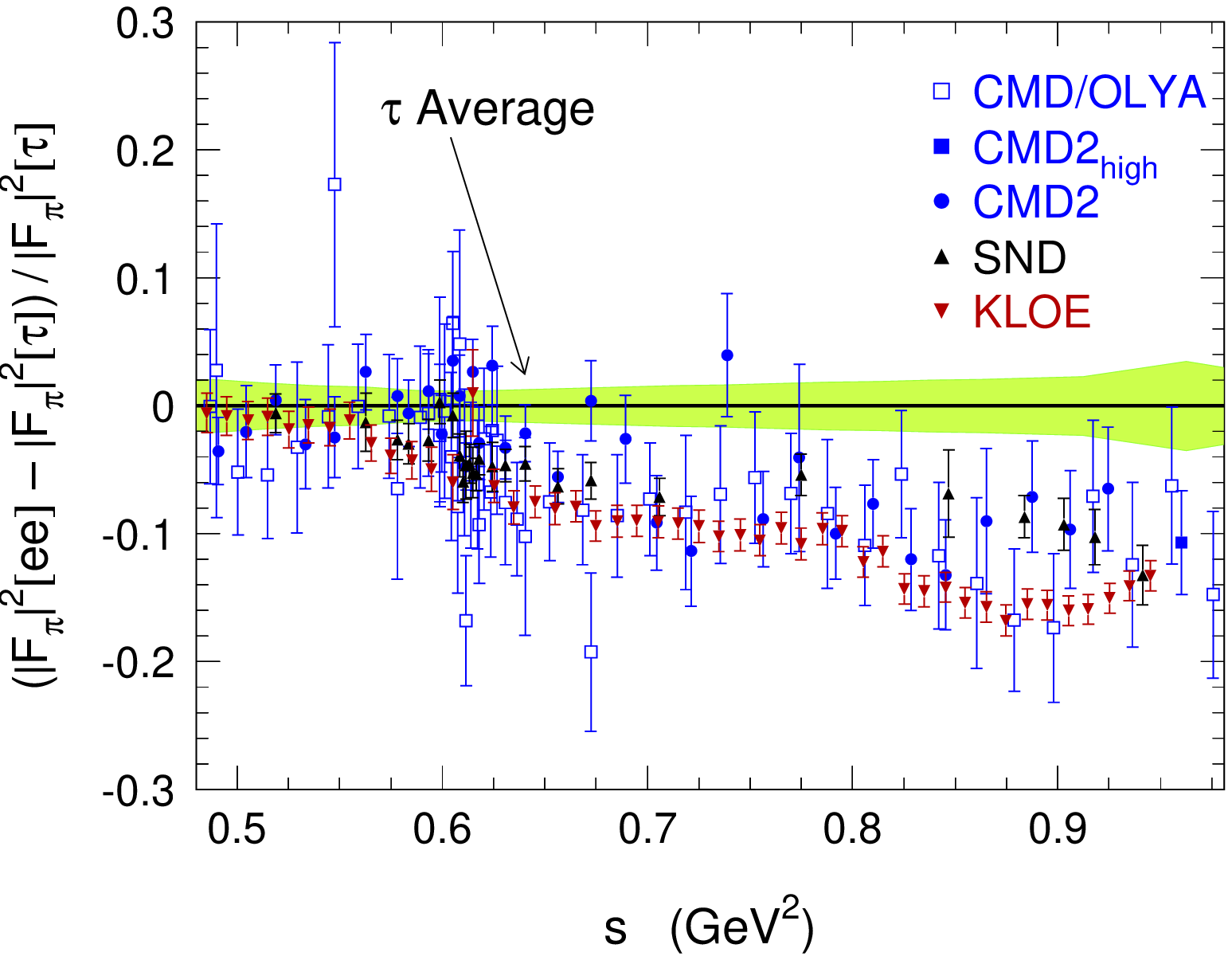,width=78mm}
             }
  \vspace{-7mm}
  \caption[.]{\label{fig:2pi_comp}
        Relative comparison of the $\pi^+\pi^-$ spectral functions
        from $e^+e^-$-annihilation data and isospin-breaking-corrected 
        $\tau$ data, expressed as a ratio to the $\tau$ spectral function.
        The shaded band indicates the errors of the $\tau$ data.
        The right hand plot emphasizes the region of the $\rho$ peak.}  
\end{figure}

A comparison between the averaged $\tau$ data and the $e^+e^-$ for the
dominant $\pi\pi$ mode is shown in Fig.~\ref{fig:2pi_comp}. 
The difference of $5-10\%$ in the energy region of $0.65-1.0\,{\rm GeV}^2$ 
is clearly visible. The difference with KLOE is even more pronounced.


\subsection{CVC}

An alternative way of comparing $\tau$ and $e^+e^-$ data is to compare
measurements of branching fractions $B$ in $\tau$ decays with 
their expectations from CVC (Conserved Vector Currect) 
using $e^+e^-$ spectral functions, duly corrected for isospin
breaking effects. The advantage of a such comparison is that the measurements
of $B$ are more robust than the spectral functions as the latter ones depend
on the experimental resolution and require a numerically delicate unfolding.  

\begin{figure}[h]
\begin{center}
\psfig{figure=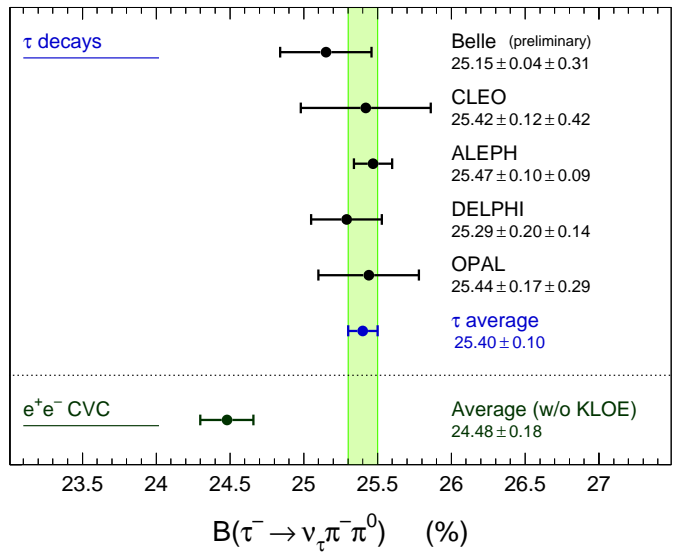,height=7.5cm}
\end{center}
\caption{The measured branching fractions for 
$\tau^-\rightarrow \nu_\tau\pi^+\pi^0$ compared to the expectation from
the $e^+e^-\rightarrow \pi^+\pi^-$ spectral function applying the 
isospin-breaking correction factors.\label{fig:cvc}}
\end{figure}

The comparison for $\pi\pi$ mode revealing a discrepancy of $4.5$ standard
deviations is shown in Fig.~\ref{fig:cvc}. Similar comparisons for decay modes $\tau^-\rightarrow \nu_\tau \pi^-3\pi^0$ 
and $\tau^-\rightarrow\nu_\tau 2\pi^-\pi^+\pi^0$ have also been made and
the differences with the corresponding $e^+e^-$ data are found respectively
at $0.7$ and $3.6$ standard deviations.

\section{Conclusion and Prospects}\label{sec:end}

The muon magnetic anomaly $a_\mu$ is one of the most precisely known 
quantities both experimentally and theoretically in the SM.
Incorporating new $e^+e^-$ data from CMD2 and SND for $\pi\pi$ mode and 
from {\sc Babar} for multihadronic modes, new SM determinations of 
$a_\mu$ have been obtained with a theoretical precision exceeding
for the first time in recent years the experimental one. 
The SM prediction is found to be smaller than the measurement by about $3.3$
standard deviations. Unfortunately one can not draw a definitive conclusion
for the moment as the $\tau$ data based prediction is in agreement with
the measurement.

Therefore it is extremely important that one clarifies the discrepancy between
the $e^+e^-$ and $\tau$ data in particular on the $\pi\pi$ mode. 
There are a number of possibilities: (1) (the normalization of) the 
$e^+e$ data is wrong, (2) the tau data are wong, (3) both are correct but 
there are unaccounted effects~\cite{mc03} which explain the discrepancy 
between the two.

Possibility (1) may be also related to the current difference (mainly on
the shape of the spectral functions) between CMD2/SND data and KLOE data
obtained respectively from the beam energy scan method and the radiative
return method. This difference is expected to be resolved soon as KLOE has
more and high quality data to be analyzed. In addition, reduced 
systematics uncertainties can be achieved if the measurement is made 
by normalizing the $\pi\pi$ data to $\mu\mu$ instead of to luminosity
using the large angle Bhabha process.

The long awaited high precision measurement in $\pi\pi$ mode from {\sc Babar}
using also the radiative return method will certainty help in clarifying
some of the issues.

On the tau side, further improvement on the high mass part of the spectral
functions is expected from large statistical data samples available at 
$B$ factories and a $\tau$-charm factory.

While the leading hadronic uncertainty gets improved with the forthcoming
high precision $e^+e^-$ (and $\tau$) data, the next item awaiting for 
significant improvement concerns the uncertainty on
the light-by-light scattering contribution $a^{\rm had,LBL}_\mu$.

Given the fact that the theoretical error is already smaller than the 
experimental one, it is timely to improve the latter. Indeed there is 
a new project BNL-E969 allowing to reduce the current error by more than 
a factor two down to $0.24\,{\rm ppm}$. We are looking forward that the
project gets funded very soon.

\section*{Acknowledgments}
The author wishes to thank the organizers of the conference for 
the invitation and M.~Davier, S.~Eidelman and A.~H\"ocker
for the fruitful collaboration.

\end{document}